\documentclass[sigconf]{acmart}
\AtBeginDocument{%
  }
    
\usepackage{cleveref}
\usepackage{enumitem}

\newcommand{\etal}[1]{{#1}~et~al.}
\newcommand{\eg}{e.g.,~}
\newcommand{\ie}{i.e.,~}
\newcommand{\textQuote}[1]{``\emph{#1}''}

\newcommand{\arbiter}{ARbiter}

\crefname{section}{Sec.}{Sec.}
\crefname{figure}{Fig.}{Fig.}
\crefname{table}{Tab.}{Tab.}

\setcopyright{none}
\acmYear{2025}
\acmDOI{}
\acmISBN{}
\acmConference[CHI'25 Workshop: Everyday AR through AI-in-the-Loop]{Second Workshop on AR and AI at CHI 2025}{April 26, 2025}{Yokohama, Japan}

\begin{document}



\title
[ARbiter]
{ARbiter: Generating Dialogue Options and Communication~Support in Augmented Reality}


\author{Juli\'{a}n M\'{e}ndez}
\orcid{0000-0003-1029-7656}
\authornote{These authors contributed equally to this research.} 
\email{julian.mendez2@tu-dresden.de}
\affiliation{%
  \institution{Interactive Media Lab Dresden, \\ TUD Dresden University of Technology}
  \city{Dresden}
  \country{Germany}
}
\author{Marc Satkowski}
\orcid{0000-0002-1952-8302}
\authornotemark[1]
\email{msatkowski@acm.org}
\affiliation{%
  \institution{Interactive Media Lab Dresden, \\ TUD Dresden University of Technology}
  \city{Dresden}
  \country{Germany}
}

\renewcommand{\shortauthors}{M\'{e}ndez and Satkowski}
\acmArticleType{Position}

\maketitle


\section{Introduction}
\label{ch:intro}

In this position paper, we propose researching the combination of Augmented Reality (AR) and Artificial Intelligence (AI) to support conversations, inspired by the interfaces of dialogue systems commonly found in videogames.
AR-capable devices are becoming more powerful and conventional in looks, as seen in head-mounted displays (HMDs) like 
    the Snapchat Spectacles~\cite{spectaclesSpectacles}, 
    the XREAL glasses~\cite{xreal}, 
    or the recently presented Meta Orion~\cite{metaOrionGlasses}. 
This development reduces possible ergonomic, appearance, and runtime concerns, thus allowing a more straightforward integration and extended use of AR in our everyday lives, both in private and at work.
At the same time, we can observe an immense surge in AI development (also at CHI~\cite{pang2025understanding}).
Recently notorious Large Language Models (LLMs) like 
    OpenAI's o3-mini ~\cite{o3mini} 
    or DeepSeek-R1\cite{guo2025deepseekr1} 
soar over their precursors in their ability to sustain conversations, provide suggestions, and handle complex topics in (almost) real time. 
In combination with natural language recognition systems, which are nowadays a standard component of smartphones and similar devices (including modern AR-HMDs), it is easy to imagine a combined system that integrates into daily conversations and provides various types of assistance. 
Such a system would enable many opportunities for research in AR+AI, which, as stated by \etal{Hirzle}~\cite{hirzle2023when}, remains scarce. 
In the following, we describe how the design of a conversational AR+AI system can learn from videogame dialogue systems, and we propose use cases and research questions that can be investigated thanks to this AR+AI combination.


\begin{figure}[t!]
    \centering
    \includegraphics[width=\columnwidth]{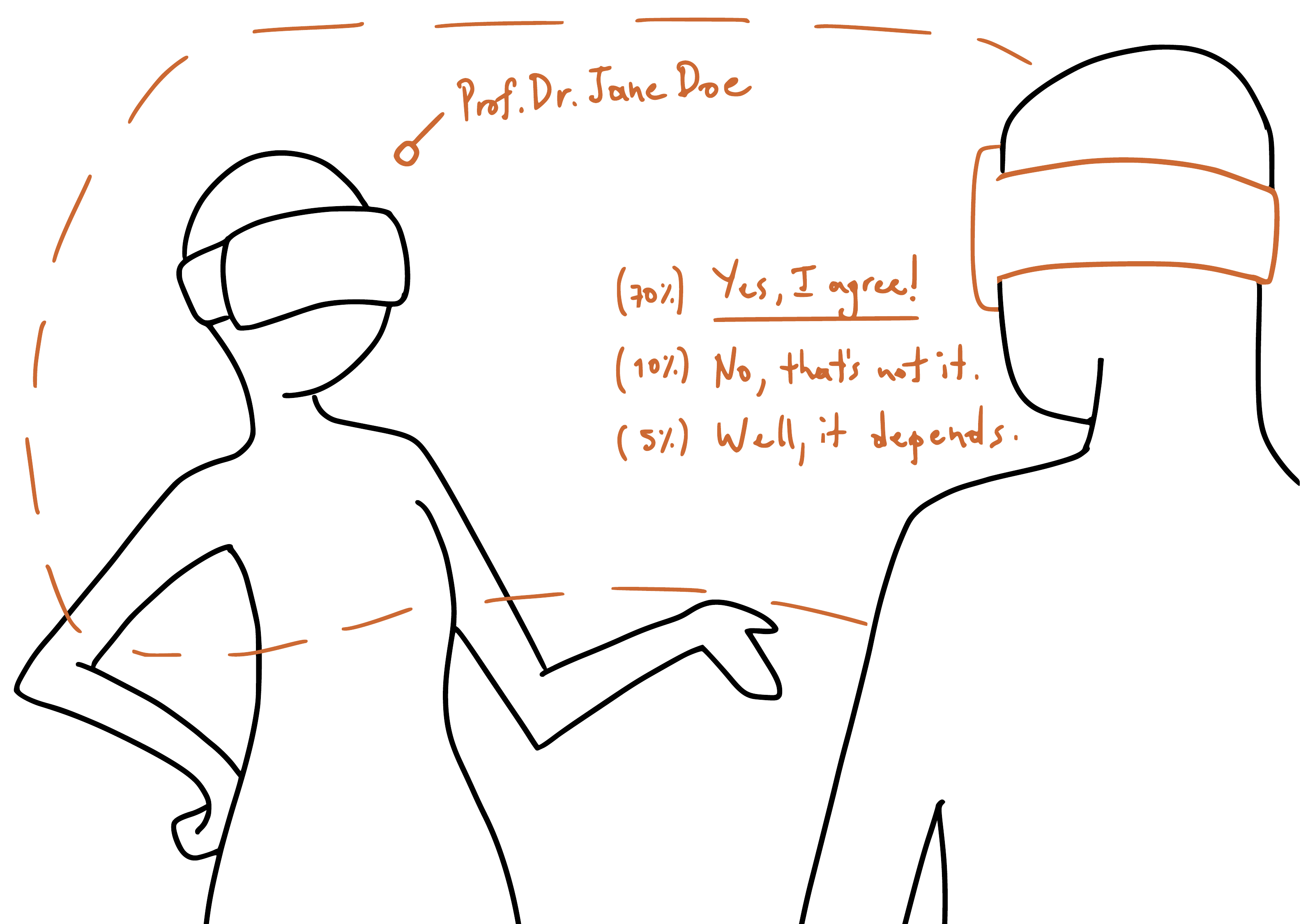}
    \caption{Sketch of our \arbiter{} concept, inspired by common user interfaces for dialogue systems in video games.}
    \Description{
    Sketch of a conversation between two users wearing head-mounted displays. An augmented reality overlay shown from the perspective of one of the users includes various elements of dialogue system user interfaces classic to video games. For instance, an indication of the name of the person that the user is talking to: Professor Doctor Jane Doe, and three dialogue options preceded by a percentage (which may relate for example to the confidence with which the artificial intelligence system suggests each option). The options read: "(70\%) Yes, I agree!", "(10\%) No, that's not it." and "(5\%) Well, it depends."  
    }
    \label{fig:concept}
\end{figure}


\section{Unintrusive, Conversational AR}
\label{ch:content}

Everyday AR -- \textQuote{AR [that is] always available to users}~\cite{suzuki2024everyday} -- must certainly consider conversations (\eg for leisure or in serious contexts). 
With LLMs revolutionizing the way users interact with knowledge, it is easy to imagine an AR+AI combination that provides individual support or plays a mediator role and steers conversations. 
In the following sections, we sketch this AI-powered arbiter (similar to the idea of \etal{Cai}~\cite{cai2024pandalens}) from an HCI perspective, as \textQuote{HCI needs to be at the center of [user interaction with LLMs]}~\cite{quere2025state}.
For that, we first describe our inspiration and concept (\cref{sec:content:concept}), highlight possible use cases and research opportunities (\cref{sec:content:useCase}), and ultimately reflect on limitations and potential problems (\cref{sec:content:limitation}).

\subsection{Inspiration \& Concept}
\label{sec:content:concept}

We imagine \textit{\arbiter{}}, an AR application that presents \textit{hints} to users during conversations. 
These hints can be unintrusively integrated into the users' field of view (see \cref{fig:concept}).
This information can be visible to some or all participants of the conversation, depending on the desired use case. 
The hints would be generated by an AI model that receives, \eg a live transcript of the spoken conversation, along with social cues collected by sensors provided by the AR HMD. 
By wrapping these inputs within a prompt that asks the AI for suggestions towards a specific outcome, it is possible to employ currently available LLMs to prototype such an application. 

Our envisioned user interface for \arbiter{} is inspired by \textit{dialogue systems} common to role-playing (\eg Mass Effect, L.A. Noir) and branching narrative games (\eg Slay the Princess, Heavy Rain).
Videogame dialogue systems give the player a predefined set of choices of what to talk about, answer, or question.
These options are hand-crafted and can affect 
\eg the game's story progression or the mood of the conversation partner.
Dialogue systems represent a simplification of the complex social communication in our daily lives, leading to a finite set of possible outcomes.
Our choices in real-life conversations also have an effect on our conversational partner and beyond, but the type of effect, to which degree, and the influence of external factors may lead to a virtually infinite set of possible outcomes.
\arbiter{} would, therefore, support users in navigating the space of possible outcomes toward a desired one.

\subsubsection*{Real-Time Conversational Support} 

Conversational user interfaces have been of interest to AI research since the 1950s~\cite{mctear2002spoken}, and recent research prototypes already integrate AI assistants into AR technology for daily tasks~\cite{cai2024pandalens, janaka2024demonstrating}. 
In fact, \etal{Cai}~\cite{cai2024pandalens} discuss how their system could, \eg help journalists during interviews by providing hints and context. 
Generally, AR devices provide a broad suite of sensors and input channels to be facilitated by an AI model.
Live data like facial recognition, body language, or conversation content can be used to create corresponding hints.
Furthermore, contextual information like the social relation between the conversation partners or specific user parameters (\eg knowledge in a given area) can be used. 
In \cref{fig:concept}, this is illustrated by showing the ``Prof. Dr.'' title in the name of Jane Doe. 
As motivated earlier, accurately predicting the outcome of a dialogue choice in real life is not always possible. 
Still, an AI could estimate the likelihood of a hint to have the desired effect on the conversation based on the recognized intent of the user. 
This is illustrated in \cref{fig:concept} with the percentages preceding the dialogue options.

\subsubsection*{UI Placement} 

While the AI produces the hints, the AR HMD handles content presentation.
A similar (starting) placement approach can be followed to those used by video games with dialogue systems. 
However, due to the dynamic nature of AR (\ie it can be used in many different environments), it is necessary to consider several factors that can impact the readability and recognizability of AR hints, like the real-world visual backgrounds~\cite{satkowski2021investigating} or the position of the conversational partner~\cite{rzayev2020effectsa}.
The positioning of UI elements may also disrupt the conversation, as it can interfere with eye contact and other social norms. Furthermore, the conversation participants can physically move, resulting in a need to adapt the placement of the AR content on the fly.
This continuous optimization problem could also benefit from AI assistance, as seen \eg in the reinforcement learning model for UI described by \etal{Todi}~\cite{todi2021adapting}.

\subsection{Use Cases \& Research Opportunities}
\label{sec:content:useCase}

The \arbiter{} concept can be applied in various scenarios,  closely related to research opportunities regarding human-AI interaction, adaptive user interface design, computer-assisted learning, collaboration, and affective computing.

\begin{description}[leftmargin=0pt,labelsep=1em]
    \item[Accessibility:] 
    Such a system can support people with communication difficulties~\cite{ciobanu2024llms} (\eg concentration problems, recognizing emotions), \eg caused by medical conditions like ADHD or autism. 
    Users of such a system could better focus on the conversation, receive reminders, and track key points of the discussion. 
    
    \item[Training:] 
    A conversational support system could be of particular interest for language learning~\cite{jeon2024systematic}, interview preparation, and public speaking rehearsing. 
    Here, hints can provide sentence formulations, translations, and assessments of performance with respect to dynamically set goals.
    
    \item[Awareness:] 
    \arbiter{} can provide important contextual information like titles (see \cref{fig:concept}) or authorities, as well as detailed information about the expertise, knowledge, cultural background, or family situation of the participants.
    
    \item[Mediation:] 
    In debates or similar situations, the participants could share visibility of the hints, allowing \arbiter{} to intervene by setting speak timers, summarizing points of interest, or helping navigate the conversation agenda.
    
    \item[Reasoning:] 
    Hints could also help to connect and form logical, sound argumentations~\cite{binyousuf2024llm} and to handle and counter possible logical flaws of the conversational partner.
    Interest in this possible use case can also be seen in upcoming workshops at CHI'25~\cite{mitEnhancedReasoning, aitoolsforthought2025Workshop} 
    
    \item[Leisure \& Self- betterment:] 
    Further use cases, such as simulating conversation scenarios, supporting creative writing, emotional regulation, etc., may also be supported by such a system. 
\end{description}

\subsection{Limitations \& Concerns}
\label{sec:content:limitation}

Besides all the research opportunities and lifestyle improvements that the \arbiter{} concept could bring, several pitfalls, limitations, and concerns must be discussed as well. 
\begin{description}[leftmargin=0pt,labelsep=1em]
    \item [LLM Hallucinations:]
    Despite the impressive state of current LLMs, these models can hallucinate~\cite{aubinlequere2024llms,cai2024pandalens}, which may lead to dangerous, unfactual information and unhelpful options. Overreliance on AI systems carries many risks exemplified by recent \textQuote{legal, academic and enterprise incidents}~\cite{aporiaRisksOverreliance}. 
    As such, special care must be taken when employing a system like \arbiter{} for sensitive topics, political debates, etc.
    
    \item [AI Biases \& Dictation:]
    Furthermore, deliberately integrating biases into or spreading misinformation through our conversation are cause for concern~\cite{quere2025state,pang2025understanding}.
    Allowing AI systems to ``dictate'' the conversations moves us towards dystopian scenarios, tied to the discourse on the long-term negative effects that technology can have on our cognitive skills and ability to establish natural connections~\cite{qawqzeh2024exploring, erosion}.
    
    \item[Security \& Privacy: ] 
    \arbiter{} would potentially process personal information of the conversation participants to, \eg  effectively query for contextual information. Thus, security concerns arise as private information may be shared with third parties (\ie as the AI ``reports'' on the participants~\cite{10.5555/3350443}). 
    This can, at minimum, hinder the conversations and, in worse cases, lead to non-compliance with laws. 
    Therefore, issues regarding consent, data collection, privacy, ethics~\cite{quere2025state,hirzle2023when}, and even secure data transfer or malicious intent need to be resolved. 

\end{description}

\section{Conclusion}
\label{ch:conclusion}
We sketched \arbiter{} as an AR+AI system that can support or steer conversations, emulating video-game dialogue systems in various real-life scenarios. 
This synthesis of ubiquitous display space and advanced AI assistance brings us closer to Sutherland's \textit{ultimate display}~\cite{sutherland1965ultimate}, another concept that, like conversational AI interfaces, surpasses 50 years of age. 
While both AI and AR technologies still need to ripen individually---with many limitations to address---we are fascinated by the possibilities that concepts like \arbiter{} enable.

\begin{acks}
This work was supported by the Deutsche Forschungsgemeinschaft (DFG, German Research Foundation) grant 389792660 as part of TRR~248 -- CPEC (see \url{https://cpec.science}).
\end{acks}

\bibliographystyle{ACM-Reference-Format}
\bibliography{bib}

\end{document}